\documentclass[preprint,showpacs,preprintnumbers,amsmath,amssymb]{revtex4}
\usepackage{graphicx,color}

\begin{document}
\title{Low magnetic field induced capacitance changes in ferronematics}
\author{Nat\'alia Toma\v{s}ovi\v{c}ov\'a$^1$,   Milan Timko$^1$,
Zuzana Mitr\'oov\'a$^1$, Martina Konerack\'a$^1$, Michal
Raj\v{n}ak$^1$, N\'andor \'Eber$^2$, Tibor
T\'oth-Katona$^2$\footnote{Corresponding author;
tothkatona.tibor@wigner.mta.hu}, Xavier Chaud$^3$, Jan Jadzyn$^4$
and Peter Kop\v{c}ansk\'y$^1$} \affiliation{$^1$Institute of
Experimental Physics, Slovak Academy of Sciences,
Watsonov\'a 47, 04001 Ko\v{s}ice, Slovakia \\
$^2$Institute for Solid State Physics and Optics, Wigner Research
Centre  for Physics,  Hungarian Academy of Sciences, H-1525
Budapest, P.O.Box 49, Hungary\\
$^3$Grenoble High Magnetic Field Laboratory, Centre National de la
Recherche Scientifique, 25 Avenue des Martyrs, Grenoble, France \\
$^4$Institute of Molecular Physics, Polish Academy of Sciences,
60179 Poznan, Poland}

\date{\today}

\begin{abstract}
The response in capacitance to low external magnetic fields (up to
0.1~T) of suspensions of spherical magnetic nanoparticles,
single-wall carbon nanotubes (SWCNT), SWCNT functionalized with
carboxyl group (SWCNT-COOH) and SWCNT functionalized with
Fe$_3$O$_4$ nanoparticles in a nematic liquid crystal has been
studied experimentally. The volume concentration of nanoparticles
was $\phi_1$~=~10$^{-4}$ and $\phi_2$~=~10$^{-3}$. Independent of
the type and the volume concentration of the nanoparticles, a
linear response to low magnetic fields (far below the magnetic
Fr\'eederiksz transition threshold) has been observed, which is
not present in the undoped nematic.
\end{abstract}

 \maketitle

\section{Introduction}

The orientational order of liquid crystals can be  controlled by
electric or magnetic fields  due to the anisotropy of dielectric
permittivity and diamagnetic susceptibility. In a restricted
geometry (between parallel substrates) this effect is called the
Fr\'eedericksz transition. The classical Fr\'eedericksz transition
in pure nematics has been studied in detail both experimentally and
theoretically {\cite{G}}. The relatively large anisotropy of the
dielectric permittivity in usual nematics triggers the orientational
response to an electric voltage already at a few volts. In contrast,
because of the small value of the anisotropy of the diamagnetic
susceptibility ($\chi_a$~$\sim$~10$^{-7}$), magnetic fields $H$ used
for the same purpose have to reach rather large values
($B=\mu_0H~\sim1$~T). In an effort to enhance the magnetic
susceptibility of liquid crystals, the idea of doping them with fine
magnetic particles was theoretically introduced by Brochard and de
Gennes. They constructed a continuum theory of magnetic suspensions
in nematic liquid crystals (ferronematics) in their fundamental
paper \cite{BG}, prior to the chemical synthesis of these systems.
Many reports have shown that doping of liquid crystals with small
amount of magnetic nanoparticles can lead to the decrease as well as
to the increase of the threshold ${B_F}$ of the magnetic
Fr\'eedericksz transition \cite{NT,PK1,Petrescu,Martinez,PK2},
depending on the anisotropy of diamagnetic susceptibility $\chi_a$
of the nematic host, and on the initial mutual orientation of the
nematic director $\mathbf n$ and the magnetic moment $\mathbf m$ of
the magnetic particles. For example, when $\chi_a>0$ and $\mathbf m
\parallel \mathbf n$, doping with magnetic nanoparticles decreases
$B_F$ \cite{NT,PK1}, while in case of combination $\chi_a<0$ and
$\mathbf m \parallel \mathbf n$ \cite{PK2}, or that of $\chi_a>0$
and $\mathbf m \perp \mathbf n$ \cite{Kopcansky2001}, doping with
the nanoparticles increases $B_F$.

In recent works by Podoliak et al. \cite{Podoliak}, and Buluy et al.
\cite{Buluy} both experimental and theoretical investigations have
been reported about the {\it optical response} of suspensions of
ferromagnetic nanoparticles in nematic liquid crystals on the
imposed magnetic field (which finally leads to a Fr\'eedericksz
transition). The authors have measured an additional, linear
response in ferronematics at low magnetic fields (far below $B_F$).
In their analysis a theoretical model based on the Burylov-Raikher
theory {\cite{Burylov1,Burylov2}} has been presented. We have to
note here that in Refs. \cite{Podoliak,Buluy} a weak bias magnetic
field ($B_{bias} \approx 2$~mT) has also been applied parallel to
$\mathbf n$ in order to align the magnetic dipole moments of the
nanoparticles parallel with $\mathbf n$ as an initial condition.
These recent works have inspired us to get more insights by
approaching the phenomenon from a different experimental aspect: to
perform {\it capacitance measurements} on the nematic liquid crystal
6CHBT doped with spherical magnetic nanoparticles, single-wall
carbon nanotubes (SWCNT), SWCNT functionalized with carboxyl groups
(SWCNT-COOH) and SWCNT functionalized with Fe$_3$O$_4$ nanoparticles
(SWCNT/Fe$_3$O$_4$), however, without an aligning bias magnetic
field ($B_{bias}=0$).

\section{Experimental}

\begin{figure}
\begin{center}
\includegraphics[width=20pc]{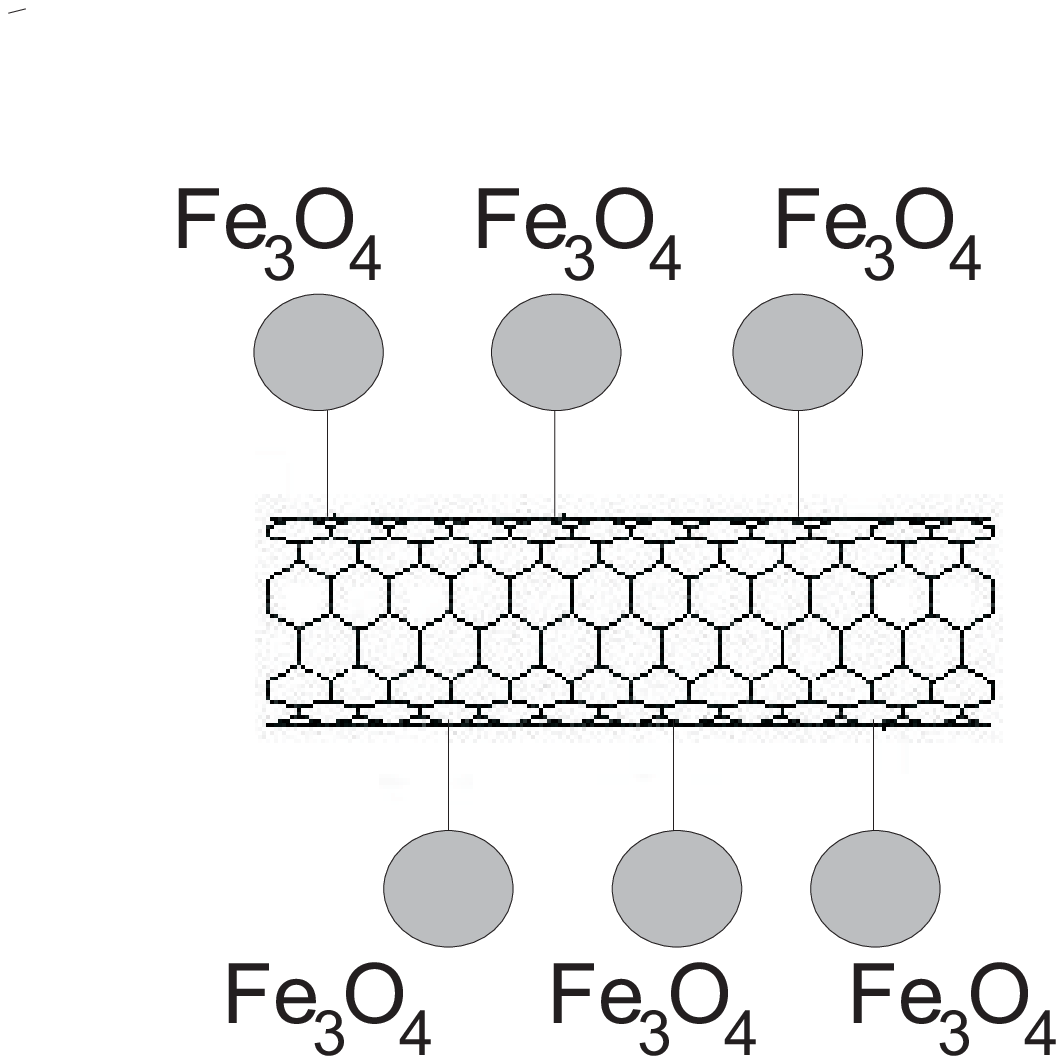}
\caption{Schematic picture of the SWCNT functionalized with
Fe$_3$O$_4$ nanoparticles (SWCNT/Fe$_3$O$_4$).} \label{sample}
\end{center}
\end{figure}

The synthesis of the spherical magnetic nanoparticles was based on
co-precipitation of Fe$^{2+}$ and Fe$^{3+}$ salts by NH$_4$OH at
60~$^\circ$C {\cite{PK1}}.
 The size and morphology of
the particles were determined by transmission electron microscopy
(TEM). The mean diameter of the obtained spherical magnetic
nanoparticles was 11.6~nm. SWCNT  and SWCNT-COOH (produced by
catalytic chemical vapor deposition technique) were purchased from
Cheap Tubes Inc., and then functionalized with Fe$_3$O$_4$
nanoparticles. The synthetic route of the functionalization is
described in details in Ref.~\ {\cite{ZM}}, and the final
SWCNT/Fe$_3$O$_4$ product is shown schematically in
Fig.~\ref{sample}. SWCNTs (of mixed chirality) exhibit in about
60$\%$ semiconducting property and in about 40$\%$ metallic one.
The length of nanotubes ranged from 0.5 $\mu$m to 2 $\mu$m, their
outer diameter from 1 nm to 2 nm and inner diameter: 0.8 nm - 1.6
nm. The magnetic properties of all prepared nanoparticles were
investigated by SQUID magnetometer (Quantum Design MPMS 5XL). The
ferronematic samples were based on the thermotropic nematic
4-(trans-4'-n-hexylcyclohexyl)-isothiocyanatobenzene (6CHBT),
which was synthetized and purified at the Institute of Chemistry,
Military Technical University, Warsaw, Poland. 6CHBT is a
low-temperature-melting enantiotropic liquid crystal with high
chemical stability {\cite{D}}. The phase transition temperature
from the isotropic liquid to the nematic phase has been found to
be 42.6~$^\circ$C. The heptan-based dispersion of spherical
magnetic nanoparticles was added to the liquid crystal in the
isotropic state under stirring until the solvent has evaporated.
The doping with nanotubes was done by adding particles under
continuous stirring to the liquid crystal in the isotropic phase.
The  volume concentrations of the nanoparticles were
$\phi_1$~=~10$^{-4}$ and $\phi_2$~=~10$^{-3}$, i.e., small enough
to keep the interparticle magnetic dipole-dipole interaction
ignorably small \cite{Burylov2}.

\begin{figure}
\begin{center}
\includegraphics[width=30pc]{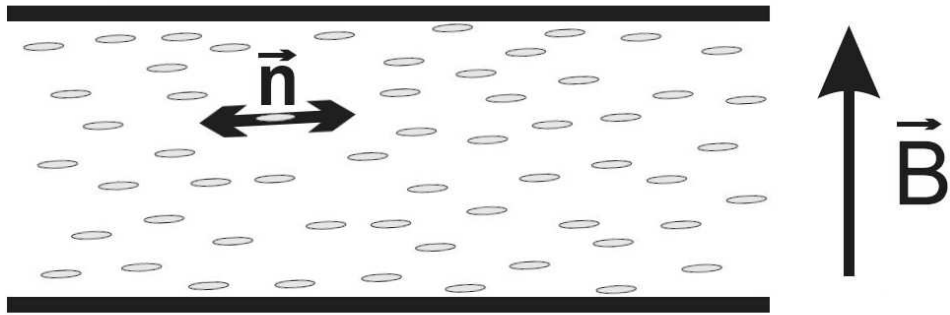}
\caption{Cross section of the cell in the initial state, which also
demonstrates a small ($\approx 3^o$) pre-tilt $\theta_0$.}
\label{schema}
\end{center}
\end{figure}

Structural transitions in ferronematic samples have been monitored
by capacitance measurements in a capacitor made of indium-tin-oxide
(ITO) coated glass electrodes. The capacitor with the electrode area
of approximately ($1 \times 1$)~cm$^2$ has been connected to a
regulated thermostat system, with a temperature stability of
0.05~$^\circ$C. Measurements were performed at the temperature of
35~$^\circ$C. The distance between the electrodes (the sample
thickness) was $D = 5$ $\mu$m. The capacitance was measured at the
frequency of 1~kHz and voltage of 0.1~V by a high precision
capacitance bridge Andeen Hagerling. In the experiments, the liquid
crystal had planar alignment. The planar alignment was achieved by
polyimide layers rubbed antiparallel, which produces a small
pre-tilt $\theta_0$ typically between $1^{\circ}$ and $3^{\circ}$ --
see e.g., in Ref.~\cite{Qi2008}. The capacitance of the cells had a
value $C_0$ for this alignment. The magnetic field was applied
perpendicular to the electrode surfaces (as shown in
Fig.~\ref{schema}). The measurements were done immediately after the
cells were filled.

\section{Results}

In the pure 6CHBT the magnetic Fr\'eedericksz transition starts at
2.63~T. Due to doping, the Fr\'eedericksz threshold is shifted to
lower values {\cite{PK1,ZM}}, but it is still higher than 2~T.

\begin{figure}
\begin{center}
\includegraphics[width=30pc]{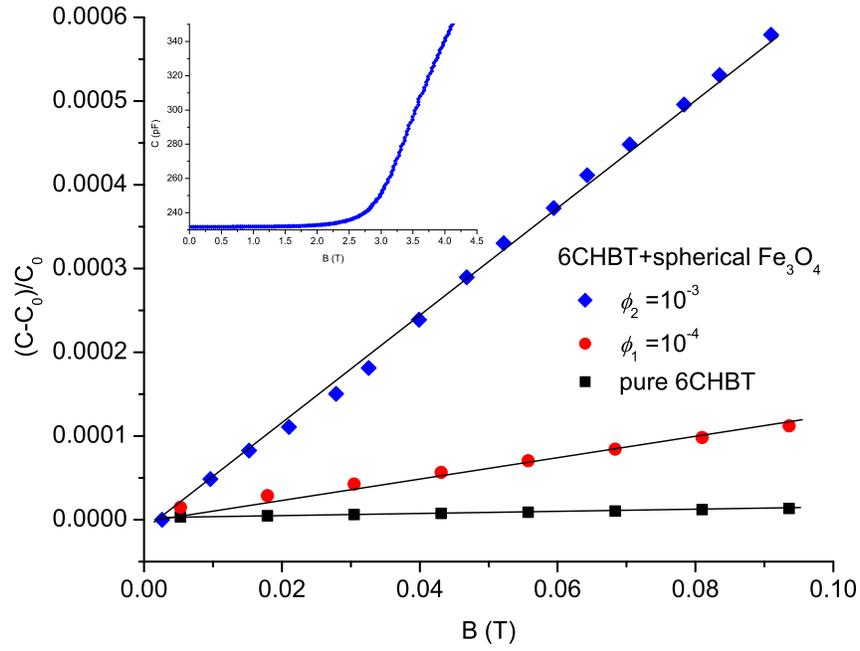}
\caption{(Color online) Relative capacitance variation vs. magnetic
field for 6CHBT doped with spherical Fe$_3$O$_4$ nanoparticles. The
inset shows the magnetic Fr\'eedericksz transition for the sample
doped with magnetic particles of volume concentration $\phi_{2}$.
Symbols are the experimental data, and the lines represent linear
fit to them.} \label{spher}
\end{center}
\end{figure}

\begin{figure}
\begin{center}
\includegraphics[width=30pc]{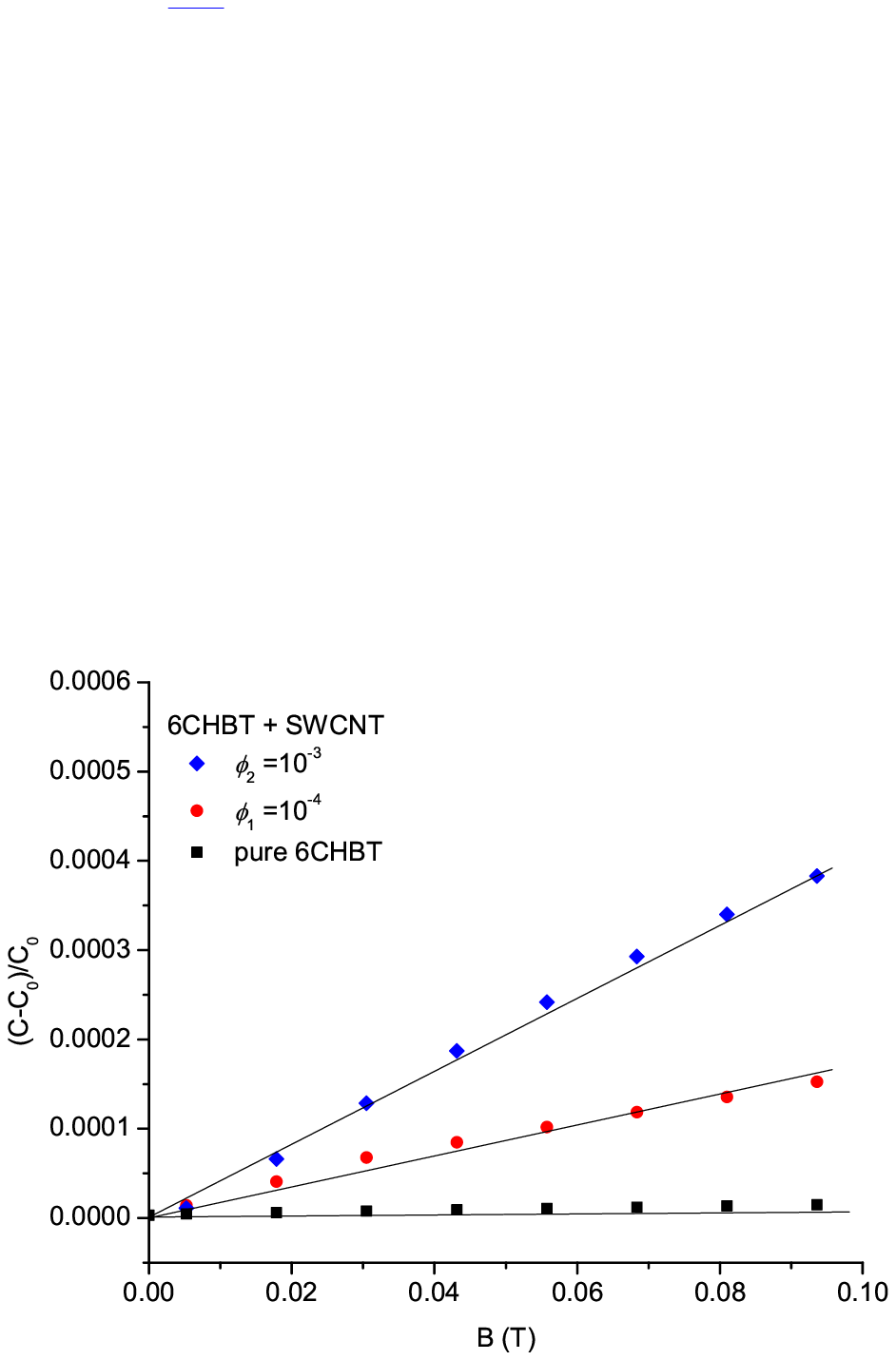}
\caption{(Color online) Relative capacitance variation vs. magnetic
field for 6CHBT doped with SWCNT. Symbols are the experimental data,
and the lines represent linear fit to them.} \label{SWCNT}
\end{center}
\end{figure}

\begin{figure}
\begin{center}
\includegraphics[width=30pc]{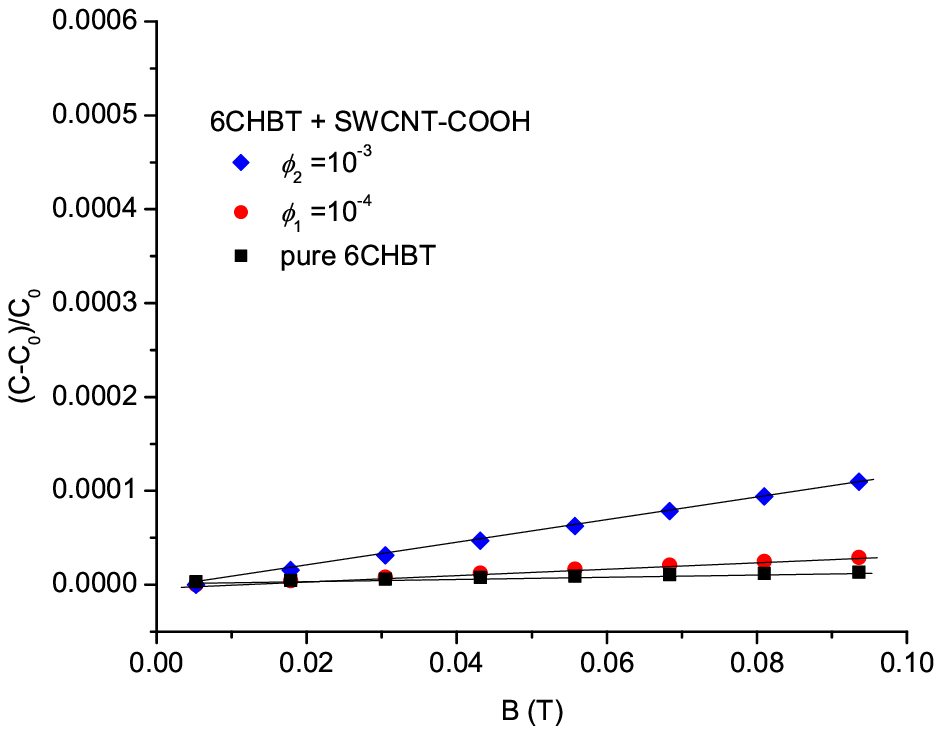}
\caption{(Color online) Relative capacitance variation vs. magnetic
field for 6CHBT doped with SWCNT-COOH. Symbols are the experimental
data, and the lines represent linear fit to them.}
\label{SWCNT-COOH}
\end{center}
\end{figure}

\begin{figure}
\begin{center}
\includegraphics[width=30pc]{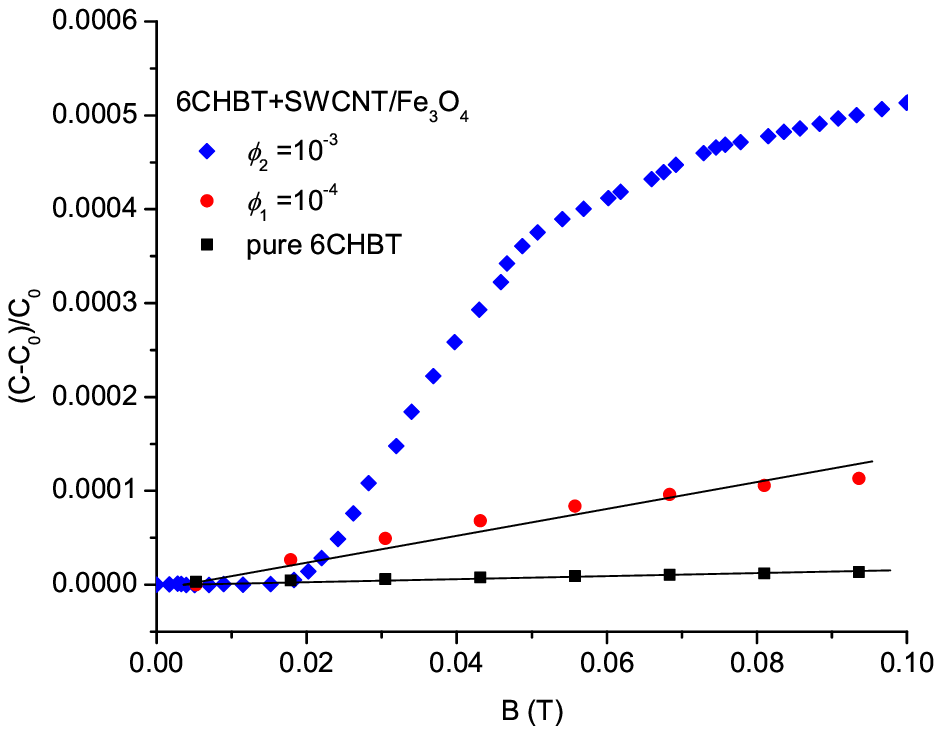}
\caption{(Color online) Relative capacitance variation vs. magnetic
field for 6CHBT doped with SWCNT/Fe$_3$O$_4$. Symbols are the
experimental data, and the lines represent linear fit to them.}
\label{SWCNT-Fe}
\end{center}
\end{figure}

Figures~\ref{spher}, \ref{SWCNT}, \ref{SWCNT-COOH} and
~\ref{SWCNT-Fe} show the relative capacitance variation $C-C_0
\over C_0$ of the 6CHBT liquid crystal doped with spherical
magnetic nanoparticles, SWCNT, SWCNT-COOH and SWCNT/Fe$_3$O$_4$,
respectively, as a function of the magnetic induction $B$ in the
low magnetic field range (up to 0.1~T), far below the threshold of
the magnetic Fr\'eedericksz transition. The figures provide a
clear evidence for a linear magnetic field dependence of the {\it
capacitance} in this low magnetic field region, i.e. for the
presence of an effect similar to that obtained by an {\it optical}
method as reported in Refs. {\cite{Podoliak,Buluy}}. The magnetic
field dependence of $C-C_0 \over C_0$ is stronger for a higher
volume concentration of nanoparticles in all cases, but it is not
present in the undoped liquid crystal. The sensitivity of the
suspensions to the field depends on the type of the nanoparticle
used. The lowest response was detected for doping with SWCNT-COOH,
while a much larger response was measured for doping with
spherical magnetic particles Fe$_3$O$_4$ and SWCNT/Fe$_3$O$_4$.
Even the nematic suspension with the net SWCNT showed a
considerably larger effect than that with SWCNT-COOH. Note that
for the suspension with the larger SWCNT/Fe$_3$O$_4$ concentration
($\phi_2=10^{-3}$) a different (nonlinear) magnetic field
dependence of the reduced capacitance is detected. A qualitatively
similar low magnetic field effect has been measured optically in
nematic colloids of carbon nanotubes filled with $\alpha$-Fe, in
which a strong aggregation has been observed on the time scale of
hours [9].

\section{Conclusions}

The data shown above confirm that ferronematic suspensions may show
well measurable response in capacitance to the magnetic field even
much below the classical magnetic Fr\'eedericksz threshold. Our
experimental results, on the one hand, are in agreement with the
recent reports \cite{Podoliak,Buluy}, regarding the dependence of
the effect on the concentration and type of nanoparticles. On the
other hand, our measurements indicate that the aligning bias
magnetic field (used in Refs.~\cite{Podoliak,Buluy}) is not an
essential prerequisite for the nematic suspensions to be sensitive
to low magnetic fields. The possible reason for this deviation
regarding B$_{bias}$ is that the ferronematics were based on
different nematic hosts in these experiments. In Ref. \cite{Buluy}
the 4-cyano-(4'-pentyl)biphenyl (5CB) has been used, while in Ref.
\cite{Podoliak} the mixture E7 served as a host, which is also
composed of various cyanobiphenyls.  For the compound 8CB (which
belongs to the same family) it was shown \cite{8CB} that it prefers
a perpendicular initial alignment ($\mathbf{n} \perp \mathbf{m}$)
between the director $\mathbf{n}$ and the magnetic moment
$\mathbf{m}$ of the nanoparticle. The role of the bias magnetic
field could be to overwrite this initial condition yielding
$\mathbf{n} \parallel \mathbf{m}$. In the 6CHBT-based ferronematics
used by us, however, $\mathbf{n}
\parallel \mathbf{m}$ is the preferred initial condition (even in
the absence of a bias field) \cite{8CB}.

A much more challenging task is to interpret the detected linear
response in capacitance to low magnetic fields. The continuum theory
allows the possibility for such a linear coupling only in the
presence of both the non-zero initial pre-tilt $\theta_0$, and the
magnetic dipole moment $\mathbf{m}$ of the nanoparticles. Namely,
without pre-tilt, in the limit of small changes, the dielectric
permittivity (and by that, the capacitance) depends only
quadratically on the change in the director out-of-plane tilt
$\Delta \theta$. However, when a pre-tilt is present, a linear term
$\theta_0 \Delta \theta$ will also enter the expression for the
permittivity. On the other hand, the free energy of the system
(which has to be minimized to obtain a solution) will contain a term
linear on the magnetic field only if the magnetic moment of the
magnetic particles $\mathbf{m}$ is taken into account through the
dipolar interaction between the particles and the magnetic field --
see Eq.~2 of Ref.~\cite{Podoliak}. The proper combination of these
effects of $\theta_0$ and $\mathbf{m}$ may eventually lead to a
solution showing the experimentally observed linear C(B) dependence.
Therefore, in this regard further theoretical and experimental
investigations are desirable, primarily concentrating on the role of
the pre-tilt $\theta_0$. Theoretically the introduction of
$\theta_0$ into the Burylov-Raikher theory, while experimentally a
systematic variation of the pre-tilt (by different surface
treatments) could serve as the first step forward.

Interestingly, 6CHBT doped with "non-magnetic" SWCNT (Fig.
\ref{SWCNT}) has a comparable low magnetic field response to those
ferronematics obtained by doping 6CHBT with magnetic particles
(spherical -- Fig.~\ref{spher}, and SWCNT/Fe$_3$O$_4$ --
Fig.~\ref{SWCNT-Fe}), while in 6CHBT doped with SWCNT-COOH the
detected response is considerably lower (though still non-zero --
see Fig.~\ref{SWCNT-COOH}). These observations lead to two main
conclusions. Firstly, both "non-magnetic" SWCNTs and SWCNT-COOH
particles have a non-zero magnetization. This has actually been
measured for our SWCNT \cite{ZM}, and is also supported by both
theoretical and experimental investigations performed on carbon
nanotubes, which have shown that nanotubes may have diamagnetic as
well as paramagnetic properties depending primarily on their
electronic structure (see e.g.,
\cite{Tsai2005,Shyu2003,Lin1995,Glenis2004,Heremans1994}). Secondly,
the functionalization (either with COOH or with Fe$_3$O$_4$) of the
carbon nanotubes (which are of mixed chirality, and exhibit both
semiconducting and metallic properties) leads to further
complications. Namely, the functionalization alters also the
interaction between the particles and the liquid crystal (i.e., the
anchoring energies are different). Therefore, magnetic properties of
the given SWCNT, as well as the type of the anchoring, and the
magnitude of the anchoring energy has to be taken into account to
interpret these results in a proper way.

The results and conclusions presented above imply that a deeper,
more detailed analysis of the theoretical model
\cite{Podoliak,Buluy}, focusing on the low magnetic field limit is
necessary for the interpretation of the low field response (both
optical and capacitive) of ferronematic suspensions. This statement
is also supported by the optical results of Ref.~\cite{Podoliak} for
higher concentrations ($\phi \geq 10^{-4}$) of nanoparticles, where
only a moderate agreement between the model and the experimental
curves has been achieved.

\section*{Acknowledgments}
The work was supported by the Slovak Academy of Sciences, in the
framework of CEX-NANOFLUID, projects VEGA 0077, the Slovak Research
and Development Agency under the contract No. APVV-0171-10, Ministry
of Education Agency for Structural Funds of EU in frame of project
6220120021, the Grenoble High Magnetic Field Laboratory (CRETA) and
the Hungarian Research Fund OTKA K81250. N.E. is grateful for the
hospitality and support provided by the Ministry of Education Agency
for Structural Funds of EU in the frame of project 26110230061.

\newpage

\newpage

\newpage

\end{document}